\documentclass[manuscript, screen]{acmart}

\renewcommand\footnotetextcopyrightpermission[1]{}
\AtBeginDocument{%
  }
    
\usepackage{tabularx}

\begin{document}

\title{What Does AI Success Look Like? A Mixed-Methods Analysis of Supranational AI Governance Instruments 
}

 \author{Delaram Golpayegani}
 \email{sgolpays@tcd.ie}
 \orcid{0000-0002-1208-186X}
 \affiliation{%
  \institution{ADAPT Centre, School of Computer Science and Statistics, Trinity College Dublin}
   \city{Dublin}
   \country{Ireland}
 }

 \author{Marta Lasek-Markey}
 \orcid{0000-0002-0183-3982}
 \email{marta.markey@tudublin.ie}
 
 \affiliation{%
   \institution{Technological University (TU) Dublin}
   \city{Dublin}
  \country{Ireland}
  }

 \author{Monique Munarini}
 \orcid{0009-0003-3966-623X}
 \email{Monique.munarini@adaptcentre.ie}
 \affiliation{%
    \institution{ADAPT Centre, Trinity College Dublin}
   \city{Dublin}
   \country{Ireland}
 }

 \author{Arjumand Younus}
 \orcid{0000-0001-7748-2050}
 \email{arjumand.younus@ucd.ie}
\affiliation{%
 \institution{University College Dublin}
 \city{Dublin}
  \country{Ireland}}

 \author{Aphra Kerr}
 \orcid{0000-0001-5445-7805}
 \email{aphra.kerr@ucd.ie}
 \affiliation{%
   \institution{ADAPT Centre, University College Dublin}
    \city{Dublin}
  \country{Ireland}}

\author{Dave Lewis}
 \email{delewis@tcd.ie}
 \orcid{0000-0002-3503-4644}
 \affiliation{%
    \institution{ADAPT Centre, School of Computer Science and Statistics, Trinity College Dublin}
   \city{Dublin}
   \country{Ireland}}

\begin{abstract}

\color{red}{This paper accepted to be published in the proceedings of ICEGOV 2026.}

\color{black} \textbf{Abstract.} In the last decade and in particular since 2018, supranational institutions have published policies, guidelines, and most recently regulations to govern development and use of Artificial Intelligence (AI). This paper aims to investigate the themes and dimensions of ``AI success'' set forth by instruments of AI governance issued by supranational bodies. To this end, we use a mixed-methods approach, combining manual thematic analysis and automated topic modelling approaches, to analyse a curated corpus of 13 documents issued by supranational bodies, including the European Union bodies, OECD, G20, UNESCO, United Nations, and the Council of Europe between 2018-2025. Based on the  analysis and through the lens of sociology of expectations, we propose a novel categorisation scheme for AI success themes emerging from supranational bodies' documents. Drawing on the findings, we advance a set of policy recommendations, setting forward the need to move beyond high-level principles toward more coordinated approaches to AI governance.

 
\end{abstract}




\maketitle

\section{Introduction}

As Artificial Intelligence (AI) becomes more prevalent in many societies, different, and sometimes conflicting visions of what constitutes \textit{successful AI} have emerged. Within these perspectives, the formal approaches defined by supranational bodies— entities that represent, or who are selected to be representative of, interests of multiple different countries— are the most influential in shaping perceptions of ``\textit{AI success}'' through establishing policies, guidelines, and rules addressing AI governance. Since 2018, in the wake of the growing public concern over harms of AI, triggered by  whistleblowing about practices in major AI companies, there has been a period of rapid growth in multi-stakeholder AI policymaking during which a number of supranational institutions have reached consensus on foundational criteria for AI governance~\cite{kerr2020expectations}. This resulted in a proliferation of political commitments and trustworthy AI guidelines across different supranational bodies, including  European Union (EU), United Nations (UN), and the Organisation for Economic Co-operation and Development (OECD). While these documents played an important role in shaping normative expectations of AI and raising awareness of AI-related risks, the non-binding nature of these instruments meant they have no direct legal obligation on organisations or states, limiting their practical effect. In response to this gap, the EU enacted the EU AI Act~\cite{eu-ai-act}, the first-ever legally binding instrument on AI establishing a risk-based approach to protect individuals' health, safety, and fundamental rights.

In this paper, we investigate the governance approaches and the potential success criteria defined in the formal documents issued by prominent supranational bodies, including the EU bodies, Council of Europe, UN, UNESCO, OECD, and G20 between 2018 to 2025, the period from early ethical and trustworthy AI guidelines to the recent EU regulation setting efforts. 
Using a mixed-methods approach that combines hand coded thematic analysis~\cite{braun2021thematic, terry2017thematic} and unsupervised topic modelling approaches, we aim to identify key \textit{themes} that signal success criteria for AI within a corpus of documents issued by the aforementioned supranational bodies. Specifically, thematic analysis is applied to a representative subset of the corpus to inductively derive contextually grounded \textit{codes}, while topic modelling is applied across all documents to identify statistically prevalent \textit{topics}. The resulting codes and topics are then interpreted through the lens of the ``Sociology of Expectations'' to generate \textit{themes}, representing broad, interpretable semantic concepts that could constitute grounds for successful AI. Through this lens, we treat these documents not merely as policy outputs, but as expressions of competing social imaginaries that seek to shape the direction of AI innovation and governance. The contributions of this paper are:
\begin{enumerate}
    \item A curated open access corpus of 13 AI documents issued by prominent supranational bodies between 2018-2025, 
    \item Combined thematic analysis and topic modelling of the documents within the corpus, yielding 157 unique concepts derived from the integration of inductively generated codes and computationally extracted topics,
     \item A multi-level categorisation of AI success themes emerging from supranational bodies, comprising four meta-themes and ten constituent themes, which are organised according to a micro-meso-macro scalar framework, 
    \item A set of policy recommendation addressing gaps in AI accountability and coordination mechanisms across the AI value chain, informed by our analysis of supranational documents. 

\end{enumerate}

The remainder of this paper is as follows: \autoref{sec:<governance>} introduces the multi-level AI Governance landscape. \autoref{sec:<related work>} reviews the related work and \autoref{sec:<methodology>} details our mixed-methods methodology combining manual thematic analysis with automated topic modelling. \autoref{sec:<results>} presents the identified codes and topics, from which the themes corresponding to AI success set forth by supranational bodies are derived in \autoref{sec:<themes>}. \autoref{sec:<policy>} provides policy recommendations and \autoref{sec:<conclusion>} concludes the paper.

\section{Multi-Level AI Governance} \label{sec:<governance>}

Governance has emerged as a key term to signal that there can be multiple forms of regulation (e.g. statutory, co-regulation and self-regulation) that operate at multiple levels, from local to
regional to global. Thus, governance in the EU can be conceptualised in vertical (state, regional, global) and horizontal terms (statutory, co-regulation and self regulation). The Oxford Handbook of AI Governance defines governance as ``the process by which societal activity is organised, coordinated, steered, and managed''~\cite[p.7]{bullock2024oxford}.

The role of AI governance is to ensure that AI systems do not undermine established legal and ethical frameworks in well established social systems. Therefore, AI governance in this research follows the notion devised in the United Nations report Governing AI for Humanity \cite{un2024governing} as any initiative aiming to address the challenges and risks posed by the development and deployment of AI systems to society, including attempts to harness AI’s potential.

The AI ecosystem involves everything related to AI, from technical developments to changes in cultural habits. To manage or even mitigate potential harmful outcomes, AI governance can be considered as any policy, regulation or process \textit{mis en place} to provide guidance, assign responsibilities and establish rights.

As we connect the notion of AI governance to responsibility, we adopt the EU definition of accountability. The EU’s Assessment List for Trustworthy AI (ALTAI), designed to operationalise the Ethics Guidelines for Trustworthy AI, defines accountability as ``the idea that one is responsible for their action, and as a corollary their consequences, and must be able to explain their aims, motivations, and reasons''\cite{eu-hleg-altai}. 
Importantly, ALTAI links accountability directly with auditability and risk management, arguing that to be accountable, organisations must establish mechanisms that ensure responsibility throughout the AI lifecycle. This includes transparent processes that can be explained to and audited by third parties, along with accessible redress mechanisms when adverse impacts occur.

\subsection{The European approach to AI Governance}

In 2018, 25 European countries signed the Declaration of cooperation on artificial intelligence (AI) \cite{eu2018_ai_cooperation_declaration} as an initiative to join efforts to regulate AI while maintaining competitiveness. In practice, at the end of the same year, the European Commission published the Coordinated plan on artificial intelligence to provide a strategic framework for national AI strategies \cite{eu2018_coordinated_plan_ai}. The European Union’s Digital Strategy \cite{ec_digital_strategy}, introduced by the European Commission in 2019 through Shaping Europe’s Digital Future, provides a roadmap for the digital transformation of sectors such as industry, public administration, and research, addressing issues including connectivity, data sharing, artificial intelligence, digital skills, and regulation. Within this framework, the 2019 Ethics Guidelines for Trustworthy Artificial Intelligence \cite{eu-hleg-trustworthyai} outline principles for responsible AI development, defining trustworthy AI as systems that are lawful, ethical, and robust from both social and technical perspectives. 

The first version of the Artificial Intelligence Act (AIA) \cite{eu-ai-act} was proposed by the European Commission in 2021. In May 2023, the final version was defined, and the legislative process, known as the trialogue, started in June 2023. The norm entered into force in August 2024. The EU chose to lay down the AI governance rules via a \textit{Regulation}, a type of secondary legislation which as per Article 288 of the Treaty on the Functioning of the European Union (TFEU) does not require any further implementing measures from the Member States. Moreover, once in force, it enjoys what is known as a direct horizontal effect, i.e., individuals may rely directly on its provisions in ordinary courts of the Member States. Unlike any other instrument of international law, EU Regulations are automatically binding not only on the member state governments, but on all persons operating on the EU single market, including private companies.

The AI Act focuses on AI governance and regulation using a risk-based approach to define, according to possible risks to society, if an AI system poses: (i) an unacceptable risk to fundamental rights, as social scoring, e.g., it should be prohibited (art 5), (ii) impact health, safety or fundamental rights of individuals, should be considered high-risk and should be heavily regulated and often monitored (art 6 to 15), (iii) or AI systems that may pose limited risks but still lower than the ones considered high-risks to citizens will still require some oversight and comply with transparency obligations (art 50) , (iv) and the ones that in general are not harmful to citizens will be considered minimal risk. The legislation also differentiates AI systems from general-purpose AI models (recital 97). The general-purpose AI models that pose systemic risks to fundamental rights should also comply with a specific set of regulations and monitoring, similar to high-risk systems (art 51 to 56). It is the first binding regulation on AI that affects different countries, and has an extraterritorial application, similar to the General Data Protection (GDPR), conceptualised by some as the ``Brussels' effect'' \cite{almada2024brussels}. According to the regulation, providers and deployers outside the European Union must comply with the AI Act if the outcomes from the AI system they create or use have results on EU citizens.

\subsection{International (Non-EU) Approaches to AI Governance}

Another governance level in our sample is represented by international supranational bodies whose activities goes beyond the EU, such as the United Nations, the OECD, the Council of Europe, and the G20.

The OECD Council Recommendation on AI \cite{oecd2019ai} was the first intergovernmental document on AI, adopted at ministerial level on the proposal of the Committee on Digital Economy Policy in 2019 and revised in 2023-2024 to accommodate the most recent AI developments. It also served as a basis for the G20 AI Recommendation \cite{g20_2019_ministerial}. The Recommendation was signed by all 38 OECD Members and eight non-members. Like many instruments of international law, the OECD Recommendation is addressed to governments and not to individual persons, even if the OECD takes inputs from other relevant stakeholders such as businesses, labour unions and civil society organisations. Furthermore, as a ‘Recommendation’, it is not legally binding, but it represents a political commitment of the adherents to make their best efforts to implement it. The G20 AI Recommendation follows the OECD AI Principles, reflecting a political commitment signalling an emerging global consensus around AI governance, yet it carries no direct legal obligation on organisations or states. 

The UNESCO’s Recommendation on the Ethics of Artificial Intelligence \cite{unesco2022recommendation} was adopted in November 2021 at UNESCO’s General Conference in Paris. Upon proclamation, UNESCO recommended that Member States apply it on a voluntary basis through legislative or other measures. It also urged Member States to engage with all relevant stakeholders, including companies, as well as bring it to the attention of national authorities, research and academic organizations involved in AI. The UNESCO Recommendation, therefore, is a commitment on the part of Member States, which has been signed by 193 countries. It is the only international document on AI that was created by reaching consensus among all recognized countries in the world. However, unlike Conventions, a UNESCO Recommendation is not legally binding, but it solidifies the political commitment to implement solutions in AI governance. Furthermore, like the majority of international law, it is addressed to governments of the signatories, and not to private actors, such as tech companies, albeit it does offer ‘ethical guidance’ to all AI actors, whether public or private. 

The Council of Europe Framework Convention on Artificial Intelligence and Human  Rights, Democracy and the Rule of Law was opened to signatures from member and non-member states in September 2024 \cite{council2024}. To date, it has been signed by the European Union and a number of other Council of Europe countries (e.g. the UK, Switzerland, Ukraine), as well as some non-members, such as the Uruguay, Canada and Japan. Importantly, it has not yet  been ratified by any of the signatories, i.e. it is non-binding as of March 2026. However, once ratified, it will  constitute the first-ever international legally binding treaty in the field of AI, with 18 countries, in addition to the European Union, having signed it. The Framework Convention is indicated to be technology-neutral, meaning that it does not focus on the technicalities of AI, but rather shifts the focus onto democracy, rule of law, and the impacts of AI.

\section{Related Work}\label{sec:<related work>}

To the best of our knowledge, no prior study has systematically analysed AI policies and guidelines issued specifically by supranational bodies over the period of 2018-2025. However, a body of related work examines AI governance documents more broadly using thematic analysis and topic modelling approaches, which are the methodologies used in this work. In the following, we provide a concise overview of prior work analysing AI policy documents that employs three broad methodological approaches: thematic analysis, topic modelling, or a combination of both. \textbf{On the Manual thematic analysis side}, early descriptive reviews~\cite{jobin2019global,hagendorff2019ethics} identified recurring ethical principles, including transparency, fairness, accountability, privacy, and human control, without formalising the process of theme identification. Subsequent work introduced greater methodological rigour: \cite{fjeld2020principled} organised these principles into eight thematic groupings mapped to operational domains, while~\cite{PALLADINO2023} employed NVivo-based thematic coding to arrange principles into nested hierarchies of overarching values and requirements. \textbf{On the automated side}, there are studies on using statistical topic modelling algorithms, such as BERTopic~\cite{grootendorst2022bertopic} and Latent Dirichlet Allocation (LDA)~\cite{blei2003lda} for analysing AI policies. Examples of using automated approaches include unsupervised topic modelling-based clustering applied to 12 national AI strategies to map government priorities~\cite{papadopoulos2020governments},  LDA-based analysis of China’s data governance policies to trace the evolution of regulatory focus over time~\cite{yang2022lda}, and structured topic models used for a comparative analysis of AI policies across China, EU and US ~\cite{wang2024developing}.  
Recently, there has been a growing uptake of \textbf{mixed-methods approaches} that combine  automated text analysis with manual thematic analysis to analyse AI governance documents. Example of using mixed-methods approaches include pairing automated analytics with manual content analysis to assess the inclusivity of ethical AI frameworks by Roche et al.\cite{roche2023ethics}, quantified analysis of manual coding of AI documents published by public sector, private organisations, and NGOs by Schiff et al. \cite{schiff2021ai}, and combination of topic modelling with manual theme extraction from national AI ethics guidelines by Saheb and Saheb \cite{saheb2024mapping}. In our earlier work~\cite{golpayegani2025uncovering}, we applied a combination of NVivo-supported thematic analysis and BERTopic-based topic modelling to analyse EU policies. The present work extends this by broadening the corpus to include policies and guidelines issued by international supranational bodies beyond the EU, enabling a comparative analysis between EU and non-EU institutions and offering a more comprehensive view of the global AI governance landscape.

\section{Methodology} \label{sec:<methodology>}

\subsection{Corpus Creation}

We defined the inclusion criteria to create the corpus as follows: formal policies, guidelines, and regulations with an explicit focus on AI, published between January 2018 and August 2025 by major supranational bodies, encompassing both binding (e.g. regulations) and non-binding (e.g. policies, recommendations, and guidelines) instruments, with no restrictions applied to document length. The resulting corpus encompasses 13 documents issued by three main groups of issuing bodies: 

\begin{enumerate}
    \item EU including trustworthy AI guidelines published by AI-HLEG (n=4)~\cite{eu-hleg-trustworthyai, eu-hleg-altai,eu_hleg-policy, eu-hleg-sectoral}  and EU AI Act~\cite{eu-ai-act} and its related guidelines~\cite{eu_prohibited, eu_genai_models, eu_gpai_code} (n=4),
    \item OECD~\cite{oecd2019ai} and G20~\cite{g20_2019_ministerial} (n=2),
    \item UNESCO~\cite{unesco2022recommendation}, UN~\cite{un2024governing}, and Council of Europe~\cite{council2024} (n=3).
\end{enumerate}

Over the half of the corpus (61\%) of the documents in the corpus were issued by EU bodies, which is due to the the relatively long history of EU initiatives on AI trustworthiness and governance. Soft instruments of governance (e.g. codes of ethics and guidelines) dominate the corpus with the EU AI Act~\cite{eu-ai-act} as the only regulatory instrument examined. Following FAIR data principles, the corpus is published as an open resource on Zenodo\footnote{\url{https://doi.org/10.5281/zenodo.19628323}}.

\subsection{Combining Thematic Analysis and Topic Modelling}

\begin{figure}[!h]
    \centering
   \fbox{ \includegraphics[width= 0.9
    \linewidth]{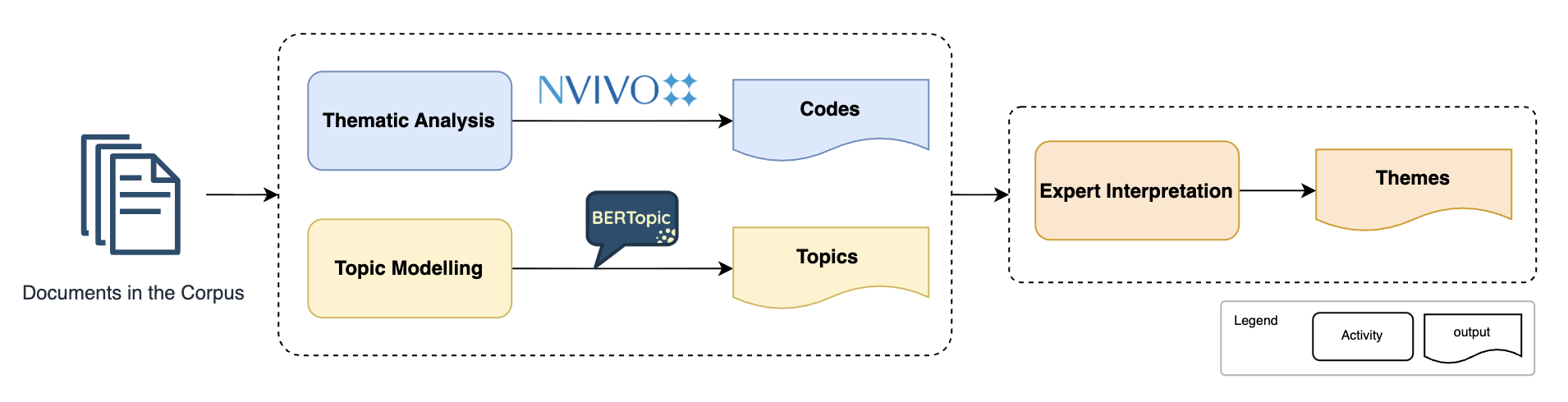}}
    \caption{Overall methodology that combines thematic analysis and topic modelling to identify themes}
    \label{fig:<methodology_overall>}
\end{figure}

As shown in \autoref{fig:<methodology_overall>}, this work employs a mixed-methods approach: it first  combines qualitative thematic analysis~\cite{Clarke2017thematic, terry2017thematic, braun2021thematic}, grounded in social sciences, with unsupervised machine learning-based topic modelling approaches to analyse the content of the documents to identify codes and topics from each document.  The integrated results of topic modelling and thematic analysis are then manually analysed by a multidisciplinary team for interpretation and identification of emerging AI success themes across the documents.

 We first applied manual thematic analysis to inductively code a representative subset of the corpus (6 out of 13), including the EU AI Act~\cite{eu-ai-act}, the HLEG's trustworthy AI guideline~\cite{eu-hleg-trustworthyai}, and documents issued by OECD~\cite{oecd2019ai}, G20~\cite{g20_2019_ministerial}, UN~\cite{un2024governing}, and UNESCO~\cite{unesco2022recommendation}. These documents were coded using the NVivo software\footnote{\url{https://nvivo.de/en/}}—a computer-assisted qualitative data analysis software in an expert-led process, performed by a legal researcher without using NVivo's AI features. The output of this process is a set of \textbf{codes}. Code refers to a concept that is \textit{manually} identified as being both present and important within the domain. 

Complementing the thematic analysis process, we used topic modelling as a computationally grounded approach to surface latent topics within documents in a scalable manner. Building upon our previous work~\cite{golpayegani2025uncovering}, we used BERTopic~\cite{grootendorst2022bertopic} as a topic modeller to identify prevalent topics within each document. In our work, \textbf{topic} refers to a concept that has been identified by a computational model, e.g. BERTopic, as a latent concept associated with a subset of sentences within each document. As topics identified by BERTopic are typically summarised by a set of salient words that tend to appear when that topic is present. Therefore, an extra expert-based step was taken to assign concise semantic labels to each topic. In this step, 2 members of the team, who are experts in law and computer science, independently labelled topics through a consensus-building process. The source code and the results are available in an archived release, linked to GitHub\footnote{\url{https://github.com/DelaramGlp/forsee_topicmodelling}}, with a persistent digital object identifier (DOI)\footnote{\url{https://doi.org/10.5281/zenodo.18303112}}, enabling reuse and reproducibility.   

Following inductive manual coding and automated topic analysis, codes and topics were interpreted through an iterative, expert-led process of interpretation to identify, refine, and name \textbf{themes}. A theme is defined as a broad interpretable semantic concept derived through expert interpretation, either from the topics generated by BERTopic or directly from the codes identified during the thematic analysis process. This interpretive process involved colour coding, clustering of related concepts, and repeated reference back to the source texts, and was conducted by a multidisciplinary team with expertise in EU law, computer science, and social science. 

Finally, informed by Sociology of Expectations, each theme was categorised using a micro-meso-macro level framework, reflecting the scale at which the expectations embedded in that theme are most appropriately addressed: by individuals or organisations (micro), at a sectoral or national policy level (meso), or at a supranational level (macro).

\section{Codes and Topics}  \label{sec:<results>}

\subsection{Codes from Manual Thematic Analysis} \label{sec:<findings_codes>}
\begin{figure}
    \centering
    \fbox{\includegraphics[width=0.9\linewidth]{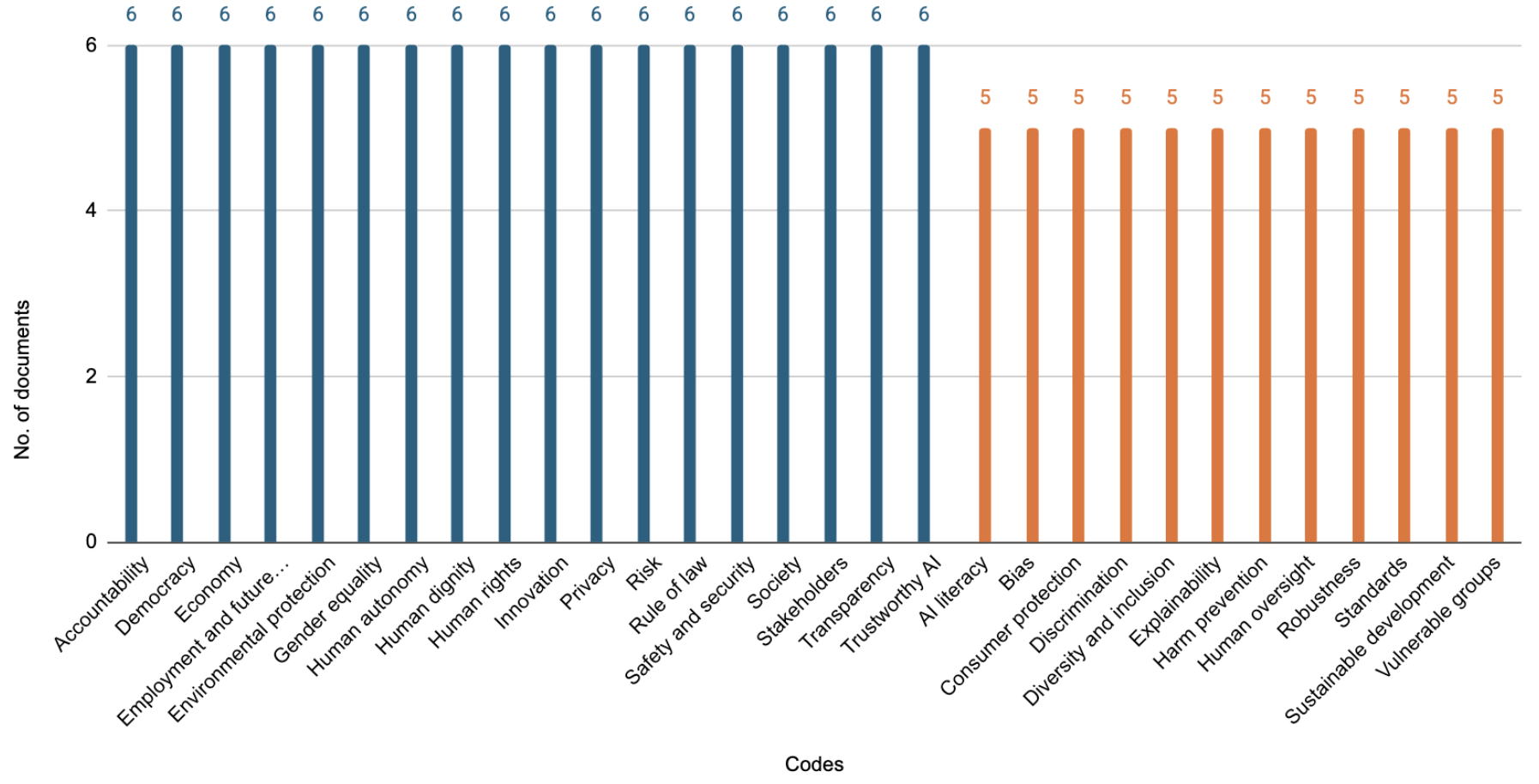}}
    \caption{Bar chart demonstrating codes appeared in five or six documents coded in the thematic analysis step}
    \label{fig:<codes-barchart>}
\end{figure}

As mentioned earlier, we manually coded  a subset of the corpus (6 documents out of 13). In total, we identified \textbf{75 unique concepts} from manual coding of these documents. Among these concepts, 18 of the concepts appeared in all 6 documents. \autoref{fig:<codes-barchart>} shows the codes that appear in all (six) or five documents that were coded manually and \autoref{tab:<codes>} presents an overview of the key codes identified in each document using manual analysis.

\begin{table}[h]
    \centering
    \small
    \caption{Summary of codes identified from thematic analysis}
    \begin{tabularx}{\textwidth}{|p{4.5cm}|p{0.7cm}| p{0.73cm}|X|}
    \hline
    \textbf{Document} & \textbf{Year }&  \textbf{No. of codes }& \textbf{Key codes identified through NVivo-supported manual thematic analysis} \\
    \hline
       HLEG Trustworthy AI Guidelines \cite{eu-hleg-trustworthyai} & 2019 & 51 & Trustworthy AI, Fundamental Rights, Ethical AI, Society, Robustness
       \\
       \hline
       OECD Recommendation of the Council on Artificial Intelligence \cite{oecd2019ai} & 2019 & 39 &  Stakeholders, Privacy, Labour and Employment, Human Rights, Risk
       \\
       \hline
       G20 Ministerial Statement on Trade and Digital Economy \cite{g20_2019_ministerial}  &2019& 45 &  Sustainable Development, Innovation, Economy, Diversity and Inclusion, Stakeholders
       \\
       \hline
       UNESCO Recommendation  on the Ethics of Artificial Intelligence \cite{unesco2022recommendation} & 2022& 49 & Human Rights, Environmental Protection, Ethical AI, Life Cycle, AI Literacy
       Gender Equality, Stakeholders, Transparency, Privacy, Diversity and Inclusion
       \\
       \hline
       Council of Europe Framework Convention \cite{council2024} & 2024 & 30 &   Human Rights, Democracy, Rule of Law, Human Oversight, Gender Equality
       \\
       \hline
       EU AI Act \cite{eu-ai-act} & 2024 & 59 & Risk, Fundamental Rights, Human Oversight, Health and Safety, Transparency, Privacy
       \\
       
       \hline

    \end{tabularx}
   
    \label{tab:<codes>}
\end{table}

\subsection{Topics from BERTopic} \label{sec:<findings_topic>}
We applied the BERTopic model to each document individually and then manually labelled the automatically generated clusters of semantically related keywords that represent topics. Across EU-issued documents, comprising  HLEG-issued guideline~\cite{eu_hleg-policy, eu-hleg-trustworthyai, eu-hleg-altai, eu-hleg-sectoral}, the AI Act~\cite{eu-ai-act}, and its related guidelines~\cite{eu_genai_models, eu_prohibited, eu_gpai_code}, we identified 100 distinct topic labels. Across these EU documents, topics evolved notably over the 2018--2025 period, shifting from topics associated with trustworthy and ethical principles in earlier documents, notably the HLEG guidelines, toward topics centred around regulatory enforcement, risk classification, and technical compliance in later ones, most markedly in the AI Act and its accompanying guidelines. 

From non-EU supranational body documents, which include~\cite{oecd2019ai, g20_2019_ministerial, unesco2022recommendation,un2024governing,council2024}, we identified 58 unique topic labels. Across non-EU documents, topics around \textit{AI and data governance}, \textit{computing infrastructure and resources}, \textit{human rights}, \textit{sustainable development}, \textit{global cooperation}, \textit{diversity and inclusion}, and \textit{AI research} are among the most common topics, reflecting the broader normative mandate of bodies such as UNESCO, the UN, and the Council of Europe. In the EU documents there has been a shift from  ethical AI in pre-AI Act documents to  operationalised legal AI with particular emphasis on regulatory enforcement in the post-AI Act era, as shown in details in our previous work~\cite{golpayegani2025uncovering}. We contend that this shift offers evidence of a broader ``regulatory turn'' in policy discourse on ethical, responsible or trustworthy AI practice and a narrowing of considerations to foreground technical and regulatory factors.

\begin{figure}[!h]
    \centering
  \fbox{\includegraphics[width=0.8\linewidth]{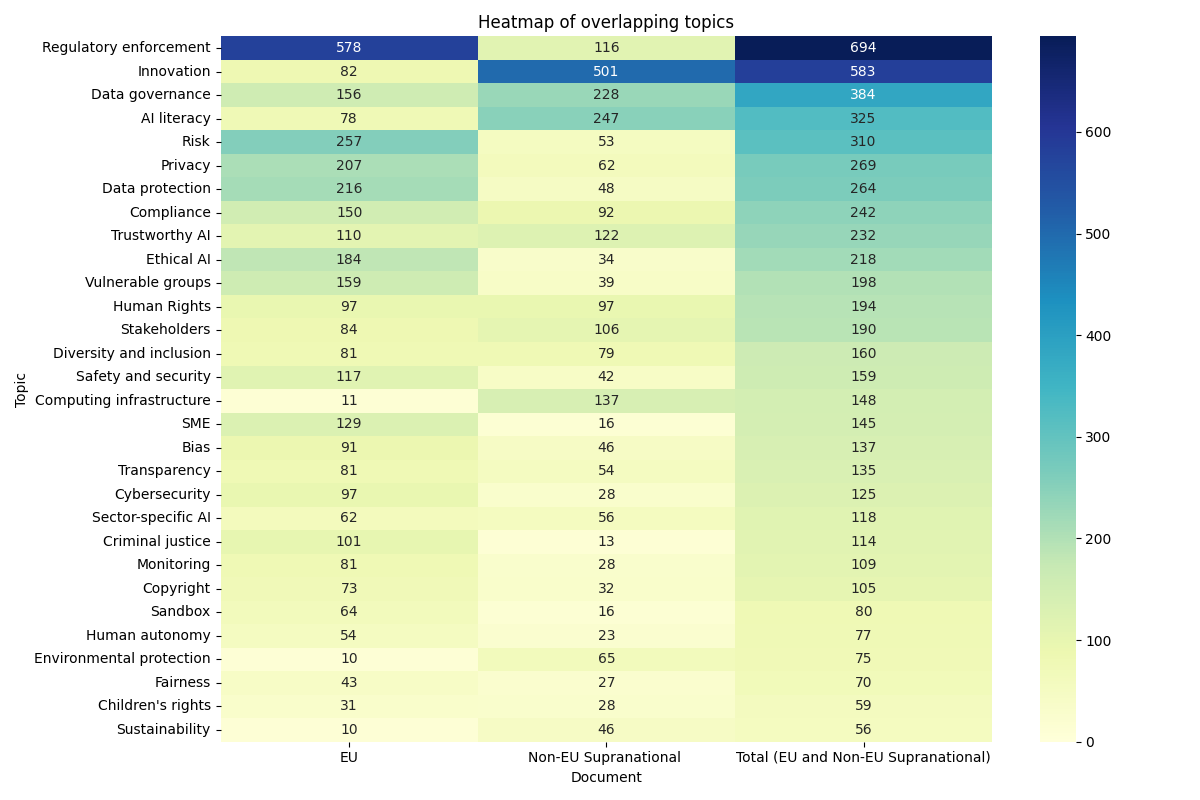}}
    \caption{Heatmap of overlapping topics between EU and non-EU supranational bodies. The numerical values displayed within the heatmap represent the aggregated frequency of that topic across each document category, based on counts calculated by BERTopic.}
    \label{fig:<bert_overlapping_heatmap>}
\end{figure}

In terms of overlaps, we identified 26 topics that are prevalent in both EU and non-EU set of documents. These overlapping topics and their frequency are illustrated using a heaptmap in \autoref{fig:<bert_overlapping_heatmap>}. Compared to the topics from the EU documents, analysis of non-EU supranational bodies revealed several distinctive emphases. Similar to the EU policies, other supranational bodies’ documents highlight the risks of AI, AI innovation, and AI literacy. Some of these documents place increasing responsibility on organisations and users to innovate and educate themselves and their users. There is also a particular attention to data governance and AI governance and stakeholders, global cooperation and collaboration to address AI challenges and opportunities. Beyond these, inclusive AI design and sustained policy dialogue are featured across the corpus. Further, sustainable development goals (SDGs), future of the environment, climate, and peace emerge as other notable topics. There are differences between the non-EU documents with the OECD, G20 and UN aligning on many topics, while the Council of Europe and UNESCO diverged, including their emphasis on vulnerable groups which they held in common with EU documents. The EU and non-EU documents align on stakeholders but AI governance as a term is more common in the non-EU documents while the EU has shifted to the language of compliance and regulatory enforcement.  By referring to both benefits and risks, supranational bodies attempt to balance positive and negative expectations compared to EU policies wherein the focus has shifted in later documents to risks. In addressing the risks, the EU policies put emphasis on technical approaches, such as risk management, and on specific high-stake applications of AI, for example for surveillance or profiling.

\section{AI Success Themes} \label{sec:<themes>}

Using cross-method comparison and integration of codes (\autoref{sec:<findings_codes>}) and topics (\autoref{sec:<findings_topic>}) and inspired by knowledge engineering and qualitative data analysis methodologies, we took an inductive approach to cluster codes and topics into meaningful categories. A \textit{theme} indicating AI success criteria was then assigned to each category, through collaborative and iterative expert interpretation. 

The integration of codes and topics revealed \textbf{157 unique concepts }across supranational body documents in the corpus. These concepts were subsequently categorised into four meta-themes: \textbf{AI technical features}, \textbf{AI Uses}, \textbf{AI Risks and Harms} and \textbf{AI Governance}. Each of these was further subdivided into constituent themes, which are also broadly categorised against a \textbf{micro-meso-macro} framing, offering a sense of where the documents expect the issues to be best governed: \textbf{Micro themes} represent expectations that are attended to by organisations or individuals, but they may do so in ways different to other organisations. \textbf{Meso themes} indicate expectations that are attended to at a specific sectoral or national policy level. \textbf{Macro themes} indicate expectations that are attended to at a supranational level and  expected to span sectoral or policy domains.

\begin{table}[!h]
    \centering
    \small
      \caption{Themes identified from supranational bodies}
    \begin{tabularx}{\textwidth}{|p{2.6cm}|X|X|X|}
    \hline
     \textbf{Meta-themes} & \textbf{Micro themes} & \textbf{Meso themes} &  \textbf{Macro themes} \\
    \hline 
       \textbf{AI Technical Issues}  &
       
       Theme 1: AI Technical Features \newline
       Theme 2: AI Quality Characteristics
      
      &
  
      &  \\

    \hline  
     \textbf{AI Use} & 
      
       &
     Theme 3: Areas of AI Use &


     Theme 4: Areas of AI Benefit \\

    \hline  
      \textbf{AI Risks and Harms} & 
      Theme 5: Risks addressed by organisation\newline
      
      &

      \multicolumn{2}{|p{6cm}|} {Theme 6: Shared Risks/Harms} 
 \\


    \hline  
     \textbf{AI Governance} & 
     Theme 7: Organisation-Level Governance Mechanism
      
      &
   Theme 8: Sectoral/Domain Oversight Mechanisms

      & 

    Theme 9: Society-Level Governance Mechanisms \newline
    Theme 10: Society-Level Governance Principles\\

    \hline  
    \end{tabularx}
  
    \label{tab:<themes>}
\end{table}

\autoref{tab:<themes>} demonstrates the themes identified from each document category. \textit{AI Technical Issues} meta-theme captures expectations directed at the technical design and quality of AI technology, encompassing two themes: \textbf{Theme 1: AI Technical Features} that consists of concepts, such as ``computing resources'' and ``model training'', that are grounded in the technical design choices made by AI developers and deployers and are understood to have a bearing on the economic and societal impact of those systems. \textbf{Theme 2: AI Quality Characteristics} that convey the types of system qualities that should be measured and controlled for AI systems to be developed and deployed in a responsible manner, such as ``transparency'', ``explainability'' and ``robustness''. The expectation for these themes is that they would be best addressed by individual organisations at the micro level.

\textit{AI Uses} meta-theme classifies concepts related to the application of AI across sectors and policy responsibility, and it is subdivided into two themes: \textbf{Theme 3: Areas AI Uses} identifies application areas of AI, such as ``healthcare'' and ``education'', and is addressed in meso level since it corresponds to definitions of sectors or governmental function, where existing systems of co-regulation and other public policy implementation already exist and therefore provide institutional anchor points for implementing AI-related policy initiatives.  \textbf{Theme 4: Areas AI Benefits} represents benefits of AI in a broad level, such as ``digital transformation'' and ``digital economy'' and is categorised at the macro level since they are broader functions of public policy, such as ``investment'' and ``economy'' that typically require a cross-government approach spanning multiple government departments and often strategic interaction with other national governments.

\textbf{AI Risks and Harms} categories concepts related to the identification of possible harms arising from AI and the management of the risk of such harms materialising. This represents the technocratic approach to governance employed when the impact and risk of harms is not well understood and therefore uncertainty is present at many levels of society on the appropriate actions to take. This meta-theme is broken down into two themes: \textbf{Theme 5: Risks addressed by organisations} contains concepts such as ``discrimination'' and ``confidentiality'' and is placed in the micro category. This is because it contains concepts where there exists some understanding of how risks represented can be assessed and as a result these can be undertaken, with oversight, by organisations with a higher degree of confidence that the risks will be effectively managed. \textbf{Theme 6: Shared AI Risks/Harms} includes concepts, such as ``democracy'' and ``human dignity'', where the level of understanding of harms and risks is not yet mature enough to delegate its management to individual organisations (micro-level) with confidence in the outcomes. Since this set of risk/harm concepts do not yet consistently fall entirely into specific sectors or government policy domains, this theme spans the meso and macro categorisations.

Finally, \textbf{AI Governance} invokes concepts that relate to modes, mechanisms, and principles that relate to how responsibilities for AI governance are undertaken at different levels. It signals the consolidation of a \textit{multi-level AI governance system}. This is the most expansive meta-theme, comprising four themes distributed across all three levels: \textbf{Theme 7: Organisation-Level Governance Mechanism} includes concepts that are needed to fulfil AI governance responsibilities, such as ``risk management'' and ``testing'' at an organisational level. \textbf{Theme 8: Sectoral/Domain Oversight Mechanisms} is a micro level theme, assembling concepts such as ``certification'', ``standards'' and ``incident reporting'' that would be employed by bodies related to specific sectors or with specific government mandates in contributing to systems of AI governance. At the macro level, \textbf{Theme 9: Society-Level Governance Mechanisms} collects concepts related to legal rights that are intended to be protected via AI governance, including concepts such as ``intellectual property'' and ``workers rights'' expressed right.\textbf{Theme 10: Society-Level Governance Principles} is also a macro theme, bringing together concepts that do not represent legally protected rights, but which have been introduced to help coordinate high level AI governance policy initiatives, e.g. ``trustworthy AI'' and ``AI literacy''.

\section{Discussion and Policy Recommendations} \label{sec:<policy>}

Our analysis of the corpus of supranational body documents reveals a  shared determination that \textbf{trustworthy AI} is achievable. There is also a consensus on the necessity of multi-stakeholders involvement in AI governance. Where the institutions and their documents differ is on the \textbf{mode of governance} (self-regulation vs. co-regulation) and the main criteria that should underpin AI governance.  Our analysis also shows a broadly shared positive expectation from supranational bodies that the benefits of \textbf{AI innovation}, in terms of economic benefits and in accelerating solutions to societal problems, can be realised when protections against variety of AI risks and potential harms are in place. Particularly in EU documents , and specifically in the AI Act, the expectation of \textbf{``innovation-with-protections''}  has emerged for AI governance, which indicates that protections can be implemented through \textit{regulatory compliance} centred around risk management, inspired by existing health and safety protections for products but extended to include fundamental rights protections. This ``innovation-with-protections'' expectation is also evident in the shift we observed in codes and topics identified in documents before and after the agreement of the AI Act, where self-regulation and concepts associated with ethical principles, are replaced by concepts associated with technical aspects of AI (e.g. data protection, cybersecurity) and operationalisation of regulation (e.g. compliance and risk management). The legacy of prior EU data protection legislation is strong, with data protection, privacy and vulnerable groups still to the fore in the EU AI regulatory documents as well as a new focus on safety. 

Our findings also suggest that AI success criteria is not universal and national and regional contexts shape the importance and the degree sensitivity to (potentially harmful) impacts of AI on rights and values such as privacy, autonomy, non-discrimination.

In the following, we propose a set of recommendations aimed at strengthening coherence and interoperability across AI governance frameworks. \textit{First}, the observed “regulatory turn” marks a shift from principle-based guidance to binding legal frameworks, particularly in the EU context. However, a gap persists between high-level values and their operationalisation, as well as between regulatory approaches across jurisdictions. Bridging this gap requires translating the ethical principles into technical standards, benchmarks, and metrology, supporting experimental governance mechanisms such as regulatory sandboxes, and prioritising interoperability to avoid regulatory fragmentation.

\textit{Second}, our analysis showed that accountability, which is a core principle for the EU’s high level expert group on AI in 2018, is less prevalent in later EU documents including the AI Act and related documents and is replaced by regulatory enforcement, AI value chains and compliance. This signals different approaches to the mode of governance, who governs AI and where responsibility lies. While the EU AI Act, clarify responsibilities across the AI value chain, gaps remain in holding some actors, such as data providers and model developers, accountable. They also remain insufficient in addressing structural power asymmetries. Large technology providers increasingly operate at a scale comparable to nation-states, concerns about enforceability and regulatory capture. Addressing this requires stronger supranational oversight, enhanced transparency obligations for big tech providers, and greater coordination among states to ensure that fundamental rights are upheld consistently across jurisdictions.

\textit{Third}, although supranational bodies frequently converge on common terminology, such as ``trustworthy'' or ``human-centric'' AI, these terms often reflect divergent normative commitments. This ``\textit{same terminology, different values}'' dynamic introduces ambiguity that undermines alignment and mutual recognition. Greater conceptual clarity, including explicit articulation of underlying values and creation of shared formal vocabularies and taxonomies using open data standards such as SKOS\footnote{Simple Knowledge Organization System is a W3C standard for representing concepts in hierarchies in a machine-readable manner (cf. \url{https://www.w3.org/2004/02/skos/})}, is therefore necessary to support interoperability and international cooperation.

\textit{Fourth}, our analysis revealed persistent fragmentation despite shared concerns across governance instruments. Limited coordination across stakeholders and institutions creates siloed activities that hinder effective responses to systemic risks to fundamental rights. Strengthening interoperability across governance frameworks, fostering sustained multi-stakeholder collaboration, and ensuring accountability across the entire AI value chain are essential to overcoming these limitations and improving AI governance coherence.

\textit{Finally}, the environmental impact of AI remains underrepresented across the corpus, particularly the recent EU AI policy documents pay much less attention to broader socio-ecological impacts such as environmental and social sustainability compared to the other supranational documents.  Integrating sustainability considerations into governance frameworks, through impact assessments, efficiency standards, and alignment with broader climate objectives, is necessary to address this critical blind spot.

Taken together, these recommendations set forward the need to move beyond high-level principles toward more coordinated approaches to AI governance, capable of addressing both existing gaps and emerging challenges.

\section{Conclusion} \label{sec:<conclusion>}

This paper examined how AI success is articulated by key supranational bodies through a mixed-method analysis of 13 major AI policies, guidelines, and regulations published between 2018-2025. The inductive analysis of the identified codes (from thematic analysis) and topics (from BERTopic) yielded in four meta-themes: AI Technical Issues, AI Uses, AI Risks and Harms and AI Governance, within which 10 distinct themes were identified. In our future work, we aim to expand the corpus to relevant national-level policies and broader regulatory frameworks with indirect AI implications, such as the GDPR and DSA, as well as technical AI standards.

\begin{acks}

This work has received funding from the European Commission's Horizon Europe Research and Innovation Programme under grant agreement No. 101177579 (FORSEE) and from Research Ireland at ADAPT, the Research Ireland Centre for AI-Driven Digital Content Technology at TCD and UCD \#13/RC/2106\_P2.

\end{acks}

\bibliographystyle{ACM-Reference-Format}
\bibliography{ref}

\end{document}